\documentclass[letterpaper, 10pt, conference]{ieeeconf}

\usepackage[utf8]{inputenc}

\usepackage{amsmath,amssymb,amsfonts}
\usepackage{algorithmic}
\usepackage{graphicx}
\usepackage{textcomp}
\usepackage{xcolor}
\usepackage{url}
\usepackage{cite}
\usepackage{tabularx}
\usepackage{caption}
\let\labelindent\relax
\usepackage{enumitem}
\usepackage{subfig}
\usepackage{multirow}
\usepackage{flushend}

\begin{document}

\title{LLM-enhanced Interactions in \\Human-Robot Collaborative Drawing with Older Adults
}

\author{Marianne Bossema$^{1,2}$, Somaya Ben Allouch$^{3,4}$, Aske Plaat$^{2}$ and Rob Saunders$^{2}$
\thanks{$^{1}$ Amsterdam University of Applied Sciences, The Netherlands {\tt\small m.bossema@hva.nl}}%
\thanks{$^{2}$ Leiden University, The Netherlands}%
\thanks{$^{3}$ Amsterdam University of Applied Sciences, The Netherlands}%
\thanks{$^{4}$ University of Amsterdam, The Netherlands}%
}

\maketitle
\begin{abstract}
The goal of this study is to identify factors that support and enhance older adults’ creative experiences in human-robot co-creativity. Because the research into the use of robots for creativity support with older adults remains underexplored, we carried out an exploratory case study. We took a participatory approach and collaborated with professional art educators to design a course ``Drawing with Robots'' for adults aged 65 and over. The course featured human-human and human-robot drawing activities with various types of
robots. We observed collaborative drawing interactions, interviewed participants on their experiences, and analyzed collected data. Findings show that participants preferred acting as curators, evaluating creative suggestions from the robot in a teacher or coach role. When we enhanced a robot with a multimodal Large Language Model (LLM), participants appreciated its spoken dialogue capabilities. They reported however, that the robot’s feedback sometimes lacked an understanding of the context, and sensitivity to their artistic goals and preferences. Our findings highlight the potential of LLM-enhanced robots to support creativity and offer future directions for advancing human-robot co-creativity with older adults.
\end{abstract}

\section{Introduction}
Engaging in creative activities can enhance cognitive health, emotional well-being, and social connections. These opportunities support lifelong learning and self-discovery, providing meaning and vitality in later life~\cite{fancourt2019evidence, gorny2022creative,tan2021being,fioranelli2023role,groot2021value}. Creativity support tools offer benefits such as enhancing cognitive function, self-expression, and well-being, that are especially valuable for older adults \cite{macritchie2023use}. Compared to screen-based tools, robots can add value through co-present, embodied, and multimodal interactions \cite{tanner2023older, breazeal2019designing}. Although robot-supported creativity has been studied with other target groups~\cite{ maher2012computational, ali2021social, rezwana2023designing}, its potential for older adults remains underexplored~\cite{bossema2023human}. As intelligent systems become part of daily life, understanding how older adults engage with them creatively is both timely and essential for inclusive Human Robot Interaction (HRI). Robot and LLM integrations further expand possibilities, enabling contextual understanding, flexible planning, and natural, conversational interaction~\cite{zhang2023large}. LLM-enhanced robots can support creative engagement without requiring technical expertise, an important affordance for older users.

\begin{figure}[t] 
\centering
\includegraphics[width=\linewidth]{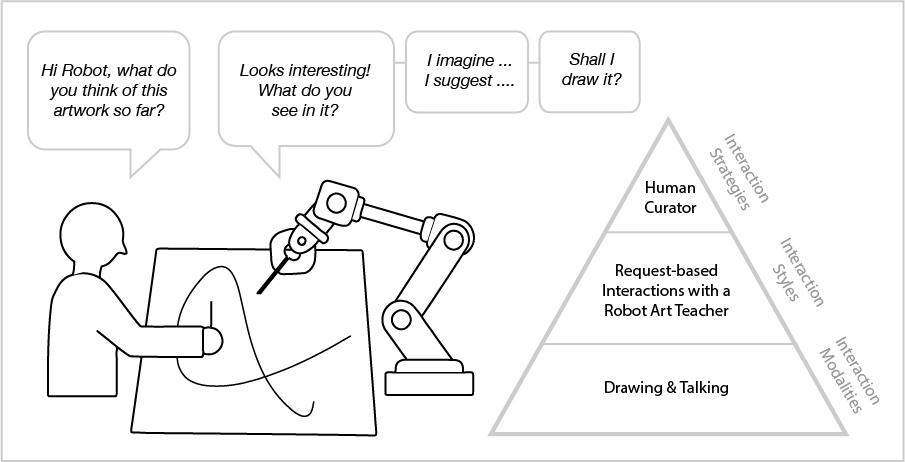} 
\caption{Impression of an interaction during a course `Drawing with Robots' for older adults, where we observed and analyzed creative collaborations using the Interaction Framework for Human–Computer Co-Creativity~\cite{kantosalo2020modalities}.}
\label{fig:visual-abstract}
\end{figure}

We conducted a structured exploratory case study using a participatory approach. The study prioritized ecological validity by observing real-world interactions within a creative course, to inform the development of well-founded hypotheses for subsequent empirical research. In line with participatory design principles, our aim was to understand how older adults engage with robots in familiar, meaningful contexts. This approach is particularly appropriate for older adults, with diverse life experiences and digital literacy~\cite{rogers2022maximizing, ostrowski2024promising, lee2023designing}. To foster engagement and gather actionable feedback, participants were actively involved in shaping the human-robot interactions. We organized an eight-week course ``Drawing with Robots'' for seniors~(n=18), together with professional art educators. Drawing offers a familiar, engaging activity for older adults, supporting social interaction, adaptive cognitive stimulation, and self-expression \cite{bone2024participatory, lin2020effects}. This setting also provided a multimodal, open-ended context for studying co-creative HRI. Robots were selected and behaviors designed based on Woolridge \& Jennings’ typology of intelligent agents~\cite{wooldridge1995intelligent}. Data analysis was guided by Kantosalo et al.’s Interaction Framework for Human-Computer Co-Creativity~\cite{kantosalo2020modalities}. This framework models co-creative interactions, where both partners contribute to a shared creative process. It distinguishes interaction strategies, styles and modalities, highlighting agency and collaborative roles, initiative, turn-taking, and communication channels.

This study provides novel insights into robot creativity support for older adults, grounded in the lived experiences of the target group. Contributions to HRI include: 1) a situated understanding of how older adults interact with robots in creative settings, 2) empirical insights into the use of LLM-enhanced co-creative robots, and 3) a research agenda on HRI design for robot creativity support with the target group.

In Section~\ref{section:background} we present foundational concepts and related works. Section~\ref{section:methodology} details the participatory methodology, the robots used, and the methods for data collection and analysis. Section~\ref{section:results} outlines the results, highlighting the main themes. In Section~\ref{section:discussion} we discuss findings, explanations, and limitations. Section~\ref{section:conclusion} concludes with a summary and proposed future directions.

\section{Background}\label{section:background}

\subsection{Layers of Interaction in Human–robot Co-creativity}
In this exploratory case study, we will compare HRI in different settings together with older adults. Comparing how co-creative systems support creativity is inherently complex due to differences in collaborative roles, turn-taking behaviors, and communication channels. The Interaction Framework for Human–Computer Co-Creativity by Kantosalo et al.~\cite{kantosalo2020modalities} offers a structured, generic lens for analysis by distinguishing between interaction strategies (the goals that drive behavior), styles (how behaviors unfold), and modalities (how information is exchanged). This layered approach is particularly suited to our study, enabling comparison across robot types and behaviors, and clarifying how different configurations shaped older adults’ creative engagement. We use the framework for a structured analysis of results, working towards an integrated view discussed in section V.

\subsubsection{Interaction Strategies}
Interaction strategies shape what a co-creative system focuses on and how it evolves, guided by what the system is trying to achieve~\cite{kantosalo2020modalities}. These goals are often reflected in the roles adopted during collaboration. Lubart~\cite{lubart2005can} identifies four robot roles, based on familiar interaction patterns: nanny, pen pal, coach, and colleague. Maher~\cite{maher2012computational} categorizes collaborative roles as supporting, enhancing, or generating creativity. Kantosalo and Toivonen~\cite{kantosalo2016modes} find overlap between these frameworks and introduce two modes: task-divided (distinct roles, one party evaluates) and alternating (mutual evaluation). A form of alternating co-creativity is proposed by Deterding et al.~\cite{deterding2017mixed} as Mixed-Initiative Creative Interfaces, where both agents dynamically shape the process. It remains underexplored, however, how goals and roles affect human-robot co-creativity for older adults.
 
\subsubsection{Interaction Styles}
Interaction styles describe how these behaviors unfold through patterns like turn-taking, requesting, or operating collaboratively ~\cite{kantosalo2020modalities}. Kantosalo et al. outline turn-taking, request-based, and operation-based styles, while Rezwana \& Maher~\cite{rezwana2023designing} analyze collaboration distinguishing participation (turn-taking vs. parallel), task distribution (shared vs. distributed), and initiative timing (planned vs. spontaneous). Specific behaviors within these interaction styles play a crucial role in shaping creative outcomes. Recent research in psychology and education sheds light on such behaviors that can foster creativity and collaboration. Han et al.~\cite{han2022perspective} found that perspective-taking feedback positively impacts self-reported creativity. Perspective-taking involves continuously assessing others' knowledge, goals, and intentions. Eyal et al.~\cite{eyal2018perspective} advocate for ‘’perspective getting,’’ a bottom-up approach in which individuals actively seek information from others, e.g., by asking questions and processing responses. Educational techniques like demonstration and scaffolding have boosted children's creativity. Robots prompting, suggesting, or demonstrating behaviors led to improved creative responses with children~\cite{elgarf2022creativebot, ali2021social}, consistent with educational theories of guided learning~\cite{vygotsky1978mind, sawyer2024explaining}. These behaviors remain largely unexamined in older adult contexts. The opportunities of perspective-taking and educational techniques have not yet been explored in human-robot co-creativity with older adults. 

\begin{figure}[t]
\centering
\subfloat[human-human drawing week 1\label{fig:human-human-collaborative-drawing-1}]{\includegraphics[width=0.49\linewidth]{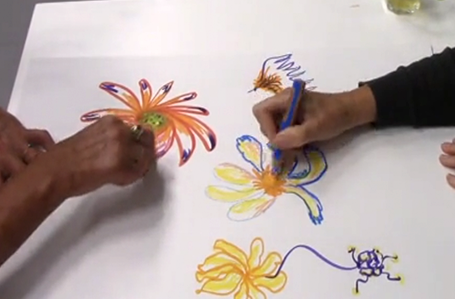}}
\hfill
\subfloat[human-human drawing week 4 \label{fig:fig:human-human-collaborative-drawing-4}]{\includegraphics[width=0.49\linewidth]{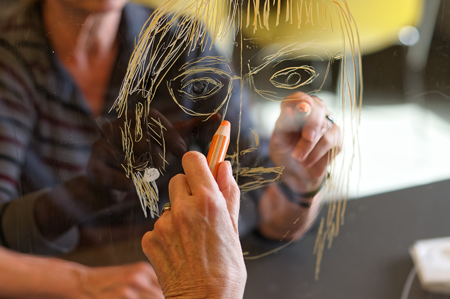}}
\\
\subfloat[presenting artwork to the group\label{fig:presenting-artwork}]{\includegraphics[width=0.49\linewidth]{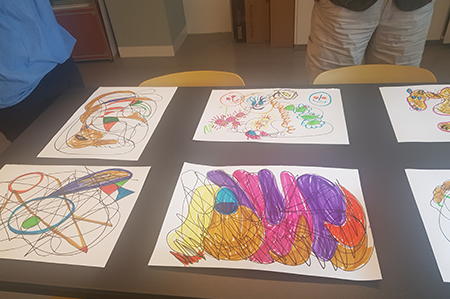}}
\hfill
\subfloat[group discussion\label{fig:group-discussion}]{\includegraphics[width=0.49\linewidth]{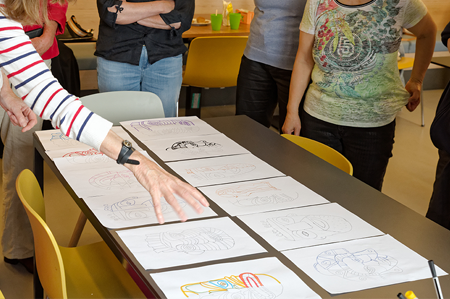}}
\caption{Human-human drawing and group discussions.}
\label{fig:drawing-activities}
\end{figure}

\begin{table*}[htbp]
\centering
\caption{Outline of the weekly course program. For image impressions see Figures 1 and 2.}
\begin{tabular}{|>{\raggedright\arraybackslash}m{0.1cm}|>{\raggedright\arraybackslash}m{4.9 cm}|>{\raggedright\arraybackslash}m{5.9 cm}|>{\raggedright\arraybackslash}m{5 cm}|}
\hline
\textbf{} & \textbf{Activity} & \textbf{Interaction Strategies, Styles, Modalities} & \textbf{Robot type/capabilities} \\
\hline
1 & Human-human co-drawing, dyadic & Peer collaboration, Dynamic, Drawing \& talking & No robots \\
\hline
2 & Designing/drawing with bristle bots, group & Robot as artist, Dynamic, Drawing & Bristle bots, reactive agents \\
\hline
3 & Drawing with mobile drawbots, group & Robot as artist, Dynamic, Drawing & Mobile bot, sensor-driven, pre-programmed \\
\hline
4 & Mirror game, human-robot, dyadic & Peer collaboration, Turn-taking, Drawing & Robot arm, sensor-driven, pre-programmed  \\
\hline
5 & Continuation game, human-robot, dyadic & Peer collaboration, Turn-taking, Drawing & Robot arm, sensor-driven, pre-programmed \\
\hline
6 & Drawing in circle, humans-robot, group & Peer collaboration, Turn-taking, Drawing and text & LLM-enhanced robot arm \\
\hline
7 & Structured dialogue, human-robot, dyadic & Robot as coach, Request-based, Drawing \& talking & LLM-enhanced robot arm, robot talking \\
\hline
8 & Open dialogue, human-robot, dyadic & Robot as coach, Request-based, Drawing \& talking & LLM-enhanced robot arm, spoken dialogue \\
\hline
\end{tabular}
\label{table:course-program}
\end{table*}

\subsubsection{Interaction Modalities}
Kantosalo et al.~\cite{kantosalo2020modalities} define interaction modalities as the mediums of communication implemented through one or more sensory channels. In human-robot collaborative drawing, communication primarily takes place through the visual channel; the drawing product. In addition, other information channels may be used. Rezwana \& Maher~\cite{rezwana2023designing} point out that most systems only support communication through the shared product. They argue that introducing diverse communication modalities can enhance coordination and improve the quality of collaboration. While LLMs offer potential for nuanced conversation, challenges in context awareness and inclusiveness remain.

\subsection{Designing HRI for Older Adults}
Building on literature about older adults’ needs, motivations, and interaction preferences is essential. Research has shown that older adults, compared to younger individuals, are less likely to use trial-and-error strategies and rely more on existing knowledge. This varies between individuals however, and depends on factors such as required effort, goals, and evaluation processes~\cite{romero2012creativity, sakaki2018curiosity, oudeyer2016intrinsic}. In addition, it has been shown that instructional frames that portray a positive account of aging can improve learning~\cite{carstensen2006aging}. Heckhausen's Motivational Theory of Lifespan Development~\cite{heckhausen2010motivational} suggests that people aim to maximize control over their environment to meet personal needs and goals. To maintain this sense of control and well-being, they adapt their motivational strategies, which in turn shapes how they explore and learn~\cite{oudeyer2016intrinsic}. This suggests that older adults' sense of autonomy and control should be taken into account when designing co-creative HRI for the target group.

\subsection{LLM-enhanced robots}
Recent advancements in LLMs open up new opportunities for conversation, supporting interaction dynamics and the negotiation of control. Zhang et al.~\cite{zhang2023large} argue that the integration of LLMs with robotics has introduced a transformative paradigm in HRI, enabling more natural dialogue, flexible interactions, and access to broad world knowledge. These advancements support more intuitive interaction, however, challenges in contextual understanding remain. Allgeuer et al.~\cite{allgeuer2024robots} explore the use of LLMs to equip robotic agents with human-like social and cognitive abilities for open-ended conversation and collaboration. They integrated an LLM with a robot's sensory perceptions and capabilities, combining speech recognition, object detection, and gesture detection. The LLM acts as the central coordinator, enabling natural and interactive control of the robot. Dogan et al.~\cite{dogan2024grace} advocate integrating human explanations with the common sense knowledge of LLMs to help robots navigate complex tasks while adhering to social norms and accommodating individual preferences. Meanwhile, Spitale et al.~\cite{spitale2024appropriateness} examine the suitability of an LLM-equipped robotic well-being coach. They emphasize the importance of follow-up questioning to prevent bias and stereotyping, arguing that the robot should avoid making assumptions without human clarification. LLMs offer new opportunities and challenges for HRI, that need further investigation with the target group of older adults.  

\subsection{Summary of Underexplored Areas}
Human-robot co-creativity with older adults has been underexplored, including the design of robot behaviors to support creativity with the population~\cite{bossema2023human}. While autonomy and (perceived) control are known to be important~\cite{romero2012creativity, heckhausen2010motivational}, it is unclear how interaction strategies, styles and modalities shape creative experiences with older adults. LLMs can enhance HRI with advanced conversational skills, but challenges in contextual understanding, personalisation and inclusiveness remain to be further explored~\cite{zhang2023large, spitale2024appropriateness}. Our study addresses these areas, looking at how we can design HRI for creativity support with older adults.

\section{Methodology}\label{section:methodology}

Between May and October 2024, we organized a course ``Drawing with Robots'' for older adults. The course consisted of eight sessions and was offered to two groups of older adults, aged 65 years and over~(n=18). For each group, an artist and a volunteer were involved, experienced in working with the target group. During the sessions, the first author of this paper attended as an observer, while a student assistant operated the robots. The project received funding from a Dutch Cultural Funding agency that aims to promote the enjoyment of life of seniors in The Netherlands through active art practice. We encouraged active engagement by using participants’ creative input as a starting point, and welcomed all feedback. The course included human-human and human-robot drawing activities (Figure~\ref{fig:weekly-activities}) with varied robot designs to observe reactions to different behaviors. We selected robots and robot behavior based on Woolridge \& Jennings’ typology of intelligent agents~\cite{wooldridge1995intelligent}, representing a continuum of designs and levels of intelligence. These ranged from reactive agents to sensor-driven and pre-programmed robots, and finally to an LLM-enhanced robot capable of spoken dialogue and image generation, simulating social and cognitive capabilities (Table~\ref{table:course-program}). This variety allowed participants to engage with different forms of human-robot collaboration and allowed for a comparative analysis of interaction strategies, styles and modalities. Data analysis was guided by the Interaction Framework for Human–Computer Co-Creativity by Kantosalo et al.~\cite{kantosalo2016modes}. This framework defines key aspects of creative human-computer collaboration at three domain-agnostic levels: interaction modalities, interaction styles, and interaction strategies, as described in Section~\ref{section:background}.

\subsection{Participants}
Participants, recruited through art teachers’ networks and an Amsterdam welfare organization, were independently living seniors aged 65 and over (16 women, 2 men). All participants previously participated in drawing courses for the elderly. The first group~(n=10), including four participants with mild cognitive impairments (MCI), attended sessions from May to mid-July 2024 in a community arts center. The second group~(n=8), with no impairments, attended from July to August 2024 on a university campus and a contemporary art center. Both groups concluded with a closing party in which participants presented their work. Informed consent was obtained from all participants; for those with MCI, consent forms were completed jointly with their informal caregivers. The study design and documents were approved by the University Ethics Review Committee.

\subsection{Collaborations}
Two experienced art educators, familiar with working with older adults, each teamed up with a trusted volunteer to lead the groups throughout the course. Preparations began with a brainstorming session with all collaborators in January 2024, followed by an April 2024 pilot test. After that we refined the robot prototypes developed by the first author, with help from computer science students. A graduate student from the University's RobotLab assisted before and during the course. After each session, the first author and artists evaluated progress and adjusted activities as needed.

\subsection{Course Design and Principles}
Participants engaged in collaborative drawing activities using physical drawing materials, chosen for familiarity and their potential to enhance creative engagement through multimodal, multi-sensory experiences~\cite{hinz2019expressive}. Physical drawing materials could also be supported by embodied, co-present human-robot interactions. Each session began with an art teacher’s introduction, including robot demonstrations when relevant. After the introduction, participants went through the same rounds: 1) an exercise to get acquainted with the materials, the robot, and the form of interaction (Figure~\ref{fig:fig:human-human-collaborative-drawing-4}); 2) in-between group reflection (Figure~\ref{fig:presenting-artwork}); 3) working towards a collaborative end product; and, 4) final group discussion (Figure~\ref{fig:group-discussion}). During the sessions, the art teachers provided feedback and answered questions. The first author observed human-human and human-robot drawing interactions, to identify factors that support creative experiences. Observations were recorded in the form of notes, photographs, video, and audio.

\begin{figure}[t]
\begin{center}
\begin{tabular}{cc}
\subfloat[a bristle bot\label{fig:a-bristle-bot}]{\includegraphics[width=0.45\linewidth]{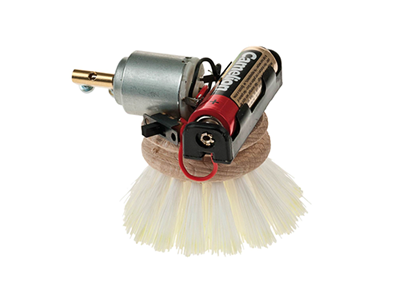}} & \multirow{-7.2}{*}{\subfloat[{uFactory Lite 6}\label{fig:ufactory-xarm-lite-6}]{\includegraphics[width=0.45\linewidth]{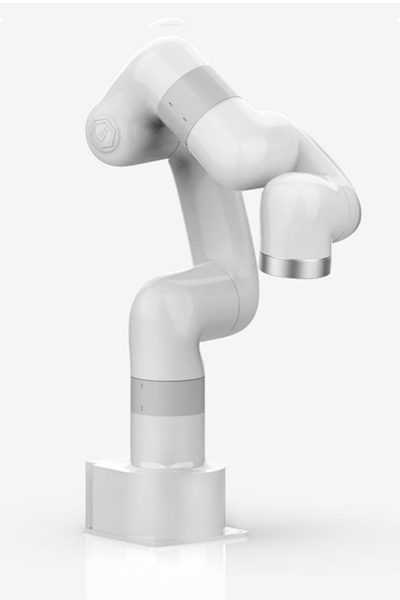}}} \\
\subfloat[{iRobot Root}\label{fig:irobot-root}]{\includegraphics[width=0.45\linewidth]{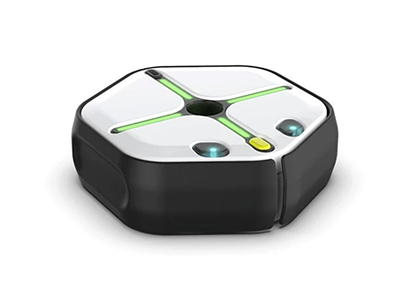}} \\
\end{tabular}
\caption{Robots used in the course.}
\label{fig:drawing-robots}
\end{center}
\end{figure}

\subsection{Course Activities}
The course activities and robot prototypes are described in the following subsections, see Table~\ref{table:course-program} for an overview. 

\subsubsection{Human-Human Collaborative Drawing}
In the first session, participants worked in pairs to create a shared collaborative drawing (Figure~\ref{fig:human-human-collaborative-drawing-1}). They were free to choose an approach and a subject. It was emphasized that the goal was to co-create an artwork and to use the entire canvas as a shared space. Participants were given watercolor pencils in vivid colors and water pens-brushes with a refillable water reservoir-allowing for blending and watercolor effects. 

\subsubsection{Human-Robot Collaborative Drawing}
The robot prototypes used in the following sessions had different designs and capabilities for collaborative behavior, see Figure~\ref{fig:drawing-robots}. We observed participants' reactions and interactions with the various robots in the following activities.

\begin{figure*}[t]
\centering
\subfloat[Week 2: Group drawing activity with bristle bots.\label{fig:group-drawing-activity-with-bristle-bots}]{\includegraphics[width=0.19\linewidth]{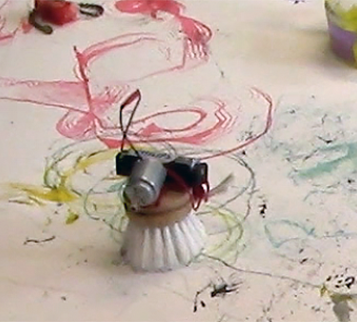}}
\hfill
\subfloat[Week 3: Group drawing activity with mobile drawbot.\label{fig:group-drawing-activity-with-mobile-robot}]{\includegraphics[width=0.19\linewidth]{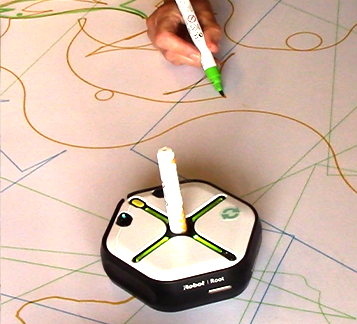}}
\hfill
\subfloat[Weeks 4-5: Dyadic imitation games with a robot arm.\label{fig:dyadic-imitation-games-with-a-robot-arm}]{\includegraphics[width=0.19\linewidth]{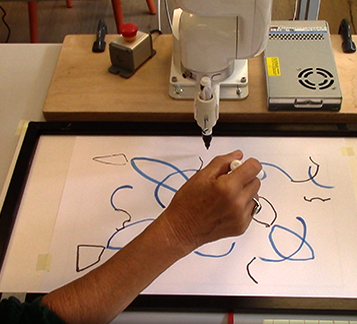}}
\hfill
\subfloat[Week 6: Group drawing activity with a robot arm.\label{fig:group-drawing-activity-with-a-robot-arm}]{\includegraphics[width=0.19\linewidth]{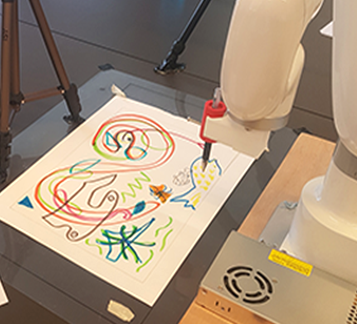}}
\hfill
\subfloat[Week 7-8: Dialogue with an LLM-enhanced robot arm.\label{fig:conversations-with-an-LLM-enhanced-robot-arm}]{\includegraphics[width=0.19\linewidth]{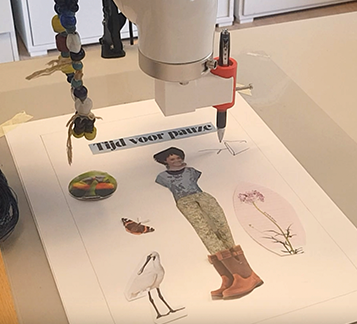}}
\caption{Weekly human-robot drawing activities.}
\label{fig:weekly-activities}
\end{figure*}

\textbf{Making small drawing machines} using bristle bots (Figure~\ref{fig:a-bristle-bot}) with DC motors. The bots could move around by vibrating their bristles. These reactive agents lacked perception, and their movements depended on motor vibrations, structure, and collisions. Their small size allowed for easy repositioning. Participants experimented with bots dragging materials dipped in paint, creating patterns. After individual trials, all bots were placed on a shared canvas for a collaborative artwork (Figure~\ref{fig:group-drawing-activity-with-bristle-bots}), with participants naming and introducing their robots.

\textbf{Drawing with a mobile robot} allowed observation of collaboration with a sensor-driven agent (Figure~\ref{fig:group-drawing-activity-with-mobile-robot}). Two iRobot Root robots, hexagonal and measuring 170 mm in diameter and 45 mm in height~\cite{iRobot}, were used (Figure~\ref{fig:irobot-root}). Equipped with color and touch sensors, they responded to canvas colors and touch input. Programmed via the iRobot® Education Python® SDK~\cite{iRobotSDK}, one robot drew curved lines and circles, while the other drew straight lines and triangles. Participants controlled the robots by repositioning them, touching them to start/stop or drive forward, and using colors on the canvas to influence drawing behavior. In the first round, groups of 4–5 participants experimented with patterns and compositions. In the second round, they chose to either continue drawing with the robot or proceed without it to elaborate on the artwork.

\textbf{Drawing games with a robot arm} were introduced in the following two weeks. We used a Ufactory Lite 6 robot arm~\cite{xArm} (Figure~\ref{fig:ufactory-xarm-lite-6}), programmed via the xArm Python SDK~\cite{xArmSDK}, with an infrared touch frame detecting user drawing movements. The activities were dyadic 'imitation games' where the robot took turns with a participant, transforming their strokes (Figure~\ref{fig:dyadic-imitation-games-with-a-robot-arm}). In the first game, the robot mirrored strokes to create masks, ornaments, or insects. In the second game, it repeatedly extended strokes from their endpoints, allowing participants to create plants or recursive patterns.

\textbf{Robot arm with a multimodal LLM} were used in the final three sessions, equipped with a webcam to capture drawings. We integrated the OpenAI Developer API~\cite{openai2024api} using a Python program. 

In session 6, the robot joined a group activity where participants sat in a circle and passed around drawings to which they added something. (Figure~\ref{fig:group-drawing-activity-with-a-robot-arm}). When receiving a drawing, the robot captured it and sent it to GPT-4o~\cite{openai2024api}, along with a list of words linked to folders containing SVG files of human-made sketches~\cite{eitz2012hdhso}. GPT-4o selected a previously unused word, and the corresponding SVG file was converted into points for the robot to draw. Location and scale were determined by identifying the largest open space using the OpenCV Distance Transform algorithm~\cite{opencv_library}. 

In session 7, we programmed the robot for structured dialogues, enabling voice output via Whisper-1~\cite{openai2024api}. Participants presented collages to the robot (Figure~\ref{fig:conversations-with-an-LLM-enhanced-robot-arm}), which sent the artwork to GPT-4o for a description and suggestions. The robot then offered to generate an image using DALL-E 3~\cite{openai2024api} and proposed adding a drawing to the collage. If participants agreed, the assistant activated the drawing process.

In session 8, we prepared the robot arm for open dialogues by incorporating Speech-to-Text capabilities~\cite{openai2024api}. Participants could present drawings to the robot and ask open questions, for example, to tell a story about the drawing, or provide specific advice on elements of interest.

\subsection{Data Collection and Analysis}
During all sessions, the human-human and human-robot drawing interactions were recorded on video, capturing the drawing process and conversations of participants from an overhead camera. We took photographs of the interactions and drawings and audio-recorded discussions. 

We analyzed audio recordings of the retrospective discussions during sessions 7 and 8. The retrospectives supported comparing the weekly activities and gaining insight in participants' self-reported experiences. In addition, we analyzed the human-human collaborative drawing videos, recorded during the first session with both groups, to investigate the creative dialogues that occured naturally between humans. For both analyses, we applied thematic coding using ATLAS.ti software~\cite{atlasti2024}. After an initial inspection of all transcribed recordings to become familiar with the content, we conducted a subsequent round of systematic coding to identify key themes. Codes were assigned to segments of video based on recurring conversational topics, behaviors, and interactions, and were iteratively refined into final themes.

To validate our findings, we used member checking with 7 of the participants and 1 volunteer, presenting key results, illustrated with with photos and quotes. We gathered feedback on accuracy and completeness, refining conclusions to genuinely reflect participants' perspectives. 

\section{Results}\label{section:results}
The analysis of discussions from sessions 7 and 8 revealed participants’ experiences with the different robot types. The human-human collaborative drawing analysis highlighted natural creative dialogues. Using the Co-Creative Framework for Interaction Design~\cite{kantosalo2020modalities}, we identified key interaction modalities, styles, and strategies in both settings. 

\subsection{Interaction Modalities: Drawing and Talking}
\subsubsection{Analysis of Retrospective Discussions} 
Conversation with the robot became possible when we equipped the robot with an LLM and spoken dialogue functionalities. During the retrospective discussions, participants highlighted various ways these conversations could support their creative process. For example, participants noted that the robot's detailed descriptions enhanced their attentive observation~(n=5), and its suggestions or stories provided inspiration~(n=5). One participant appreciated the motivation it offered: \textit{``I really like that I don't have to make something beautiful. That I can just make something crazy. And that the robot is also happy about it and says: oh, how nice!''} Another participant valued the structured feedback: \textit{``The robot adds creativity because it clearly structures what the subject is, what you can do with it, and how to work it out.''} Conversations were seen as useful at various moments during the process~(n=2), especially when participants needed direction: \textit{``It’s inspiring when you have no idea how to proceed, but I also like to try things myself first. It could be helpful throughout the process, depending on the needs of the person drawing.''} Participants mentioned that conversation also provides opportunities for step-by-step instructions~(n=2) or personalized tasks to challenge creativity~(n=3): \textit{``Yes, giving assignments, like we got from the teacher now. That would be nice.''}

\subsubsection{Analysis of Human-Human Collaborative Drawing}
\begin{figure*}[tb]
\centering
\includegraphics[width=\textwidth]{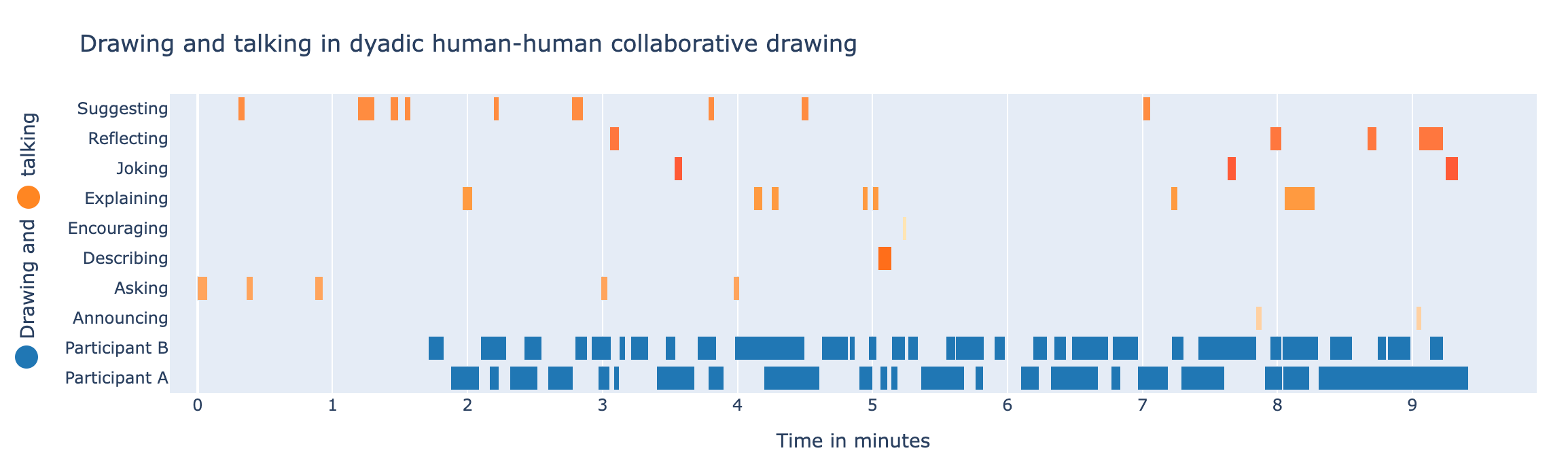}
\caption{Timeline of drawing and talking actions during a typical session of dyadic human-human collaborative drawing.}
\label{fig:drawing-and-conversation-actions}
\end{figure*}

Analysis of human-human collaborative drawing interactions revealed that all pairs naturally engaged in spoken dialogue while drawing. Throughout the interaction, participants continuously exchanged information about the drawing process and product. Conversation varied based on individual needs at different stages during collaboration. Figure~\ref{fig:drawing-and-conversation-actions} shows a timeline of drawing and conversational actions, observed during a typical session. Initially, conversation was used to negotiate goals and approach. After that, it flowed seamlessly alongside the drawing activity, allowing participants to monitor the canvas while communicating, without needing to pause or switch contexts. Participants typically started by taking turns, but would sometimes draw simultaneously, once they could anticipate each other's actions. The timeline plot (Figure~\ref{fig:drawing-and-conversation-actions}) highlights conversational actions for coordination and sense-making that occurred during the drawing process across all teams with varying frequencies and timings.

To summarize, in human-human collaborative drawing, participants naturally used spoken dialogue to build common ground. They also appreciated this communication strategy when collaborating with a robot. Conversation allowed for an extra communication channel, in addition to the drawing product itself.

\subsection{Interaction Styles: Dialogues supporting co-creativity}
Through conversation, the LLM-enhanced robot was able to make suggestions, tell stories and generate images in response to participants' artistic expressions. Recognizable analogies and semantic relationships helped people connect visual elements from their drawings to the robot's responses. Participants noted that human-robot conversation contributed to this alignment~(n=7): \textit{``The robot gives a suggestion that fits your drawing. That's good.''} Participants explained however, that the robot also missed aspects that they themselves imagined~(n=5) : \textit{``The robot mentioned it all, but didn't really go into the symbolism.''} Another participant had an abstract idea of what her drawing was about: \textit{``So for me it was speed versus frozen time, so to speak.''}, while the robot only responded to the literal things detected in the drawing. Using an LLM allowed the robot to generate answers and simulate understanding, based on the information embedded in the LLM training data. Participants' reactions to the human-robot conversations reflect that their goals, preferences, and skills were not yet sufficiently taken into account. 

The robot's drawing capabilities were not equally important to everyone. While some participants preferred drawing by themselves and were more interested in the robot's conversational capabilities for creativity support (n=5), others wanted the robot to teach them how to draw (n=2). A robot demonstrating and teaching drawing techniques seemed useful to them: \textit{``For example, I can't draw depth at all. So I can ask the robot to teach me that step by step.''} For a robot contributing to a drawing, participants felt it should align with their personal goal and style of the artwork~(n=7).

\subsection{Interaction Strategies: Roles, Dynamics and Control}
In comparison to the LLM-enhanced robot, participants found the other robots difficult to collaborate with. There were mixed feelings about the level of control they had over the creative process, illustrated by statements such as: \textit{``That little robot was really quirky''}. Participants explained that interacting with those robots left them frustrated due to their lack of control over the creative process~(n=4).

One participant explained that the LLM-enhanced robot with spoken dialogue support was the best fit for learning: \textit{``That robot gave me good examples and I could learn from that''}. Equipped with conversational skills, the robot took on an advisory role, and people consulted the robot when they wanted. These request-based interactions differed from previous sessions that involved shared canvas drawing, simultaneous work, or turn-taking. It shifted participants to a curator role, evaluating the robot's suggestions and guiding the creative process. Figure~\ref{fig:weekly-activities} gives an impression of weekly human-robot drawing activities during the course, which involved different interaction strategies, styles, and modalities. When asked about the robot as a peer collaborator taking turns, a participant responded: \textit{``Yes, that was fun, but then it's just a fun experiment. It's not like I'm making something for myself''}. The participant felt peer collaboration limited personal creative goals and development. While discussing the robot as an assistant, a participant noted it could help those unable to perform tasks themselves, a different target group than the course participants.


\section{Discussion}\label{section:discussion}
This section analyzes results, relates findings to literature, and highlights research limitations and future opportunities.

\subsection{Methodological Insights}
We investigated human-robot collaborative creativity through an exploratory, participatory approach aligned with calls for responsible, human-centered robotics~\cite{lee2023designing, ostrowski2024promising}. The eight-week course allowed participants to develop familiarity with the technology and reflect on different robot behaviors over time. Our method supports prior work advocating for sustained involvement of older adults in technology design. Member checking with participants and the instructor added depth and validity to the analysis.
Using the Interaction Framework for Human–Computer Co-Creativity~\cite{kantosalo2020modalities}, we examined how interaction strategies, styles, and modalities shaped the collaborative experience. While these dimensions were analyzed separately for clarity, they dynamically interacted in practice. For instance, spoken dialogue (modality) enabled request-based feedback (style), supporting a curator-like interaction strategy. The integration of a LLM was central to this alignment, enabling conversational flexibility that bridged modalities, styles and strategies.

\subsection{Comparing Human-human and Human-robot Drawing} While both supported co-creation, the interaction dynamics were different. Human-human drawing happened spontaneously, with shifting leader roles, while interaction styles such as turn-taking emerged naturally and changed over time. Drawing and talking were fluidly integrated, and sometimes overlapping. Human-robot collaboration relied more on structured, predefined patterns. Although the LLM-enhanced robot allowed for conversation, enabling negotiation and flexible interactions, participants in our study typically engaged in request-based exchanges, taking on a more directive role. In this situation, drawing and talking remained loosely connected. There is an opportunity to further investigate the integration of conversation in human-robot co-creative processes, and how this can support and enhance creativity for older adults.

\subsection{Layers of Interaction in Human-robot Drawing}
Participants preferred the role of human curator, using robots as creative coaches or art teachers. Through request-based interactions, they maintained control over the process, consistent with task-divided co-creativity~\cite{kantosalo2016modes}. Participants in our study were less inclined toward alternating modes, or mixed-initiative co-creativity as proposed by Deterding~\cite{deterding2017mixed}. This may be related to Heckhausen's theory on adapting strategies to maintain a sense of control and well-being ~\cite{heckhausen2010motivational}, indicating a preference for more familiar and predictable forms of collaboration. We argue that older adults' sense of autonomy and control should be taken into account by offering adaptive interaction styles and strategies.
Our findings align with prior research showing the value of diverse communication channels in creative collaboration~\cite{rezwana2023designing}, particularly spoken feedback. Educational techniques such as demonstration and step-by-step guidance were also effective~\cite{ali2021social, elgarf2022creativebot}. However, participants noted a lack of sensitivity in the robot’s feedback to their personal goals and preferences. Perspective-taking approaches~\cite{han2022perspective} could support better user modeling and context awareness, allowing robots to better tailor their responses to individual artistic goals and preferences.

\subsection{Limitations and Opportunities} 
This study has several limitations to be considered when interpreting the results. Its exploratory nature, the small gender-biased sample, and the specific settings, limit the generalizability of the findings.
In addition, we investigated first encounters with the technology, which makes a novelty effect inevitable. 
While first encounters are informative~\cite{smedegaard2019reframing}, this might have affected participants' preference for familiar collaborative roles and interaction patterns. Since familiarity develops over time, it is crucial to study how engagement and adaptation evolve as older adults become accustomed to new forms of creative collaboration. This process may also influence their preferences regarding the distribution of user and robot initiative.
Forms of verbal communication during collaborative drawing sessions have been mapped out, however, the role of nonverbal communication, and the dynamics of drawing and talking are yet to be explored. Opportunities exist to investigate the interplay of different forms of communication and how they can support the co-creative process.

\section{Conclusion}\label{section:conclusion}
We investigated interaction layers that support human-robot co-creativity for older adults. We found that in our exploratory, participatory study: a) participants preferred the role of human curator, evaluating creative suggestions from the robot as a teacher or coach; b) they favored guiding the creative process through request-based interactions; c) spoken dialogue with the robot was well received; however, d) while the robot provided inspiration through conversation, its feedback often lacked sensitivity to participants' individual artistic goals and preferences. Combining these findings with opportunities from the Discussion, we propose the following future directions.

\begin{itemize}[leftmargin=0.1in]
\item \textbf{Dynamic Interaction Modes} - Consider older adults' sense of control and autonomy when designing for the target group. As this may vary and change over time, further investigate flexible interaction modes, based on user preferences and engagement levels;
\item \textbf{Perspective-taking} - Further explore how human-robot conversation can contribute to context awareness and user modeling. Perspective-taking techniques allow for better tailoring robot responses to individual artistic goals and preferences; 
\item \textbf{Multimodal Communication} - Spoken dialogue can improve creative collaboration. Investigate how different forms of communication- i.e. verbal, nonverbal, through the creative product- can be integrated to enhance human-robot co-creative processes.
\end{itemize}

\section*{Acknowledgment}
This publication is part of the project `Social robotics and generative AI to support and enhance creative experiences for older adults', financed by Doctoral Grant for Teachers project number 023.019.021 from the Dutch Research Council (NWO). The course `Drawing with Robots' was funded by Lang Leve Kunst Fonds in The Netherlands.

\bibliographystyle{ieeetr}
\bibliography{articles}

\end{document}